\documentclass[11pt,a4paper]{article}
\usepackage{amsmath,amssymb}
\usepackage{epsfig,graphicx}

\newcommand{\ep}{\varepsilon}
\newcommand{\Li}[2]{{\mbox{Li}}_{#1}\left(#2\right)}

\begin{document}
\abovedisplayskip=0.1cm
\belowdisplayskip=0.1cm

\title{\bf 
Hypergeometric functions, their $\ep$ expansions and Feynman diagrams
}
\author{M.~Yu.~Kalmykov$^{a,b}$,
B.A.~Kniehl$^{a}$,
B.F.L.~Ward$^c$,
S.A.~Yost$^d$ 
\footnote{{\bf e-mail}: 
kalmykov@theor.jinr.ru, 
kniehl@mail.desy.de,
BFL{\underline\ }Ward@baylor.edu, 
scott.yost@citadel.edu 
} 
\\
$^a$ 
\small{\em {II.}\ Institut f\"ur Theoretische Physik, Universit\"at Hamburg}, \\
\small{\em Luruper Chaussee 149, 22761 Hamburg, Germany} \\
$^b$ 
\small{\em Joint Institute for Nuclear Research, $141980$ Dubna (Moscow Region), Russia} \\
$^c$ 
\small{\em Department of Physics, Baylor University,
One Bear Place, Waco, TX 76798, USA} \\
$^d$ 
\small{\em Department of Physics, The Citadel, 171 Moultrie St.,
Charleston, SC 29409, USA} 
}
\date{}
\maketitle

\begin{abstract}
We review the hypergeometric function approach to Feynman diagrams.
Special consideration is given to the construction of the Laurent expansion. 
As an illustration, we describe a collection of
physically important one-loop vertex diagrams for which this approach
is useful. 
\end{abstract}
\bigskip
\noindent{\bf 1. Introduction. }
Recent interest in the mathematical structure of Feynman diagrams 
has been inspired by the persistently increasing 
accuracy of high-energy experiments and the advent of the LHC epoch.
For stable numerical evaluation of diagrams, a knowledge of their analytical 
properties is necessary. We will review some of the progress in this area, 
focusing on the hypergeometric function representation of Feynman diagrams. 

Forty-five years ago, Regge proposed \cite{regge} 
that any Feynman diagram can be understood in terms of a special class 
of hypergeometric functions satisfying some system of differential equations
so that the singularity surface of the relevant 
hypergeometric function coincides with the surface of the Landau singularities 
\cite{landau} of the original Feynman diagram.\footnote{ 
For a review of different approaches to the analysis of the singularities 
of Feynman diagrams see Ref.\ \cite{book}.}
Based on Regge's conjecture, explicit systems of differential equations 
for particular types of diagrams have been constructed. For some 
examples, the hypergeometric representation for $N$-point
one-loop diagrams has been derived in Ref.\ \cite{1loop:series} via a 
series representation (Appell functions and Lauricella functions 
appear here), the system of differential equations and its 
solution in terms of Lappo-Danilevsky functions \cite{Lappo} has 
been constructed in Ref.\ \cite{1loop:diff}, and the monodromy 
structure of some Feynman diagrams has been studied in Ref.\ \cite{monodromy}.

A review of results derived up to the mid-1970's can be found in 
Ref.\ \cite{golubeva}.  It was known at that time that each 
Feynman diagram is a function of the ``Nilsson class.''
This means that the Feynman diagram is a multivalued analytical function in 
complex projective space ${{\mathbb C}{\mathbb P}^n}$. 
The singularities of this function are described by Landau's equation.
Later, Kashiwara and Kawai showed \cite{KK} that any 
regularized Feynman integral satisfies some holonomic system of 
linear differential equations whose characteristic variety is confined 
to the extended Landau variety.  

The modern technology for evaluating Feynman diagrams is based 
mainly on techniques which do not explicitly 
use properties of hypergeometric functions, but are based on relationships
among the Feynman diagrams derived from their internal structure.\footnote{By 
``internal structure,'' we mean any representation described in standard 
textbooks, such as Ref.\ \cite{QFT}.}  It was shown, for example, that there 
are algebraic relations between dimensionally regularized \cite{dimreg} 
Feynman diagrams with different powers of propagator \cite{ibp}.
Tarasov showed in 1996 that similar algebraic relations could also be 
found relating different dimensions of the
integral \cite{tarasov}. The Davydychev-Tarasov algorithm \cite{tarasov,Davydychev:1991} 
allows a Feynman diagram with arbitrary numerator  to be transformed into a linear combination 
of diagrams of the original type with shifted powers of propagators and space-time dimension,  
multiplied
by a linear combination of tensors constructed from the metric tensor and 
external momenta.  This set of algebraic relations is analogous to 
{\it contiguous relations} for hypergeometric functions.\footnote{The full set
 of contiguous relations for generalized hypergeometric functions ${}_pF_q$ 
is found in Ref.\ \cite{rainville}.}

Solving the algebraic 
relations among Feynman diagrams allows them to be expressed in terms of a 
restricted set called ``master integrals.'' Such a solution is completely 
equivalent to the differential reduction of hypergeometric 
functions \cite{D,MKL06,Tarasov}.  The technique of describing Feynman diagrams by 
a system of differential equations was further extended in Ref.\ \cite{DE},
where it was realized that the solution of the recurrence relations can 
be used to close the system of differential equations for any Feynman diagram.
This led to useful techniques for evaluating
diagrams \cite{DE:old,DE:recent}.  Most of the progress to date in this type of
analysis has been for diagrams 
related to the ``Fuchs'' type of differential equation, with three
regular singular points \cite{ADE}\footnote{The 
analysis of some diagrams with four regular singularities was done 
recently in Ref.\ \cite{heun}.}. 

Since Feynman diagrams are often UV- or IR-divergent, it is important to 
also consider the construction of the
Laurent expansion of dimensionally-regularized diagrams about integral values 
of the dimension (typically $d=4-2\ep$). This is called an
``$\ep$ expansion'' of the diagram.  For practical applications, we need 
the numerical values of the coefficients of this expansion. 
Purely numerical approaches are under development ({\it e.g.} Ref.~\cite{num}), but this is a 
complex problem for many realistic diagrams  having UV and IR singularities and several mass scales. 

The case of one-loop Feynman diagrams has been studied the most. The
hypergeometric representations for N-point one-loop diagrams with arbitrary powers 
of propagators and an arbitrary space-time dimension have been derived for 
non-exceptional kinematics\footnote{``Non-exceptional kinematics'' refers to 
the case where all masses and momenta are non-zero and not proportional 
to each other.} by Davydychev in 1991 \cite{davydychev}. His approach 
is based on the Mellin-Barnes technique \cite{BD}.  The results are expressible
in terms of hypergeometric functions with one less variable than the 
number of kinematic invariants.  

An alternative hypergeometric representation 
for one-loop diagrams has been derived recently in Ref.\ \cite{FJT03}, 
using a difference equation in the space-time dimension. 
This approach has been applied only 
to a set of master integrals\footnote{The 
hypergeometric representations of one-loop master integrals of propagator and 
vertex type have been constructed in \cite{BD,1loop:vertex}.}, 
but, fortunately, an arbitrary $N$-point function can be reduced 
to the set of master integrals analytically \cite{one-loop:reduction,one-loop:nonexceptional}.
In Ref.\ \cite{FJT03}, the one-loop $N$-point function was shown to
be expressible in terms of hypergeometric functions
of $N\!-\!1$ variables. One remarkable feature of the derived results is a
one-to-one correspondence between arguments of the hypergeometric functions and 
Gram and Cayley determinants, which are two of the main characteristics of 
diagrams.

Beyond one loop, a general hypergeometric representation is available only for 
sunset-type diagrams with 
arbitrary kinematics \cite{lauricella}, with a  simpler representation for 
particular kinematics \cite{sunset,KK08a}. 
In all other cases beyond one loop, master integrals have been expressed 
in terms of hypergeometric functions of type ${}_pF_{p-1}$ \cite{bateman}.

The program of constructing the analytical coefficients of the 
$\ep$-expansion is a more complicated matter.
The finite parts of one-loop diagrams in $d=4$ dimension are 
expressible in terms of the Spence dilogarithm function
\cite{one-loop:finite}.  However, only partial results for higher-order 
terms in the $\ep$-expansion are known at one loop. 
The all-order $\ep$-expansion of the one-loop propagator with an arbitrary 
values of masses and external momentum has been constructed \cite{DK} in 
terms of Nielsen polylogarithms \cite{Lewin}.
The term linear in $\ep$ for the one-loop vertex diagram with
non-exceptional kinematics has also been constructed in terms of Nielsen 
polylogarithms \cite{one-loop:linear}. It was shown in Ref.\ \cite{1loop:all} 
that the all-order $\ep$ expansion for the one-loop vertex with 
non-exceptional kinematics is expressible in terms of multiple polylogarithms 
of two variables \cite{Goncharov}.

Beyond these examples, the situation is less complete. The 
term linear in $\ep$ for the box diagram is still under construction. 
Some cases for particular masses\footnote{In Ref.\ \cite{FJT03}, 
box diagrams have been written in terms of the Lauricella-Saran function $F_S$
of three variables,  and a one-fold integral representation 
was established for their all-order $\ep$ expansion. 
However, it is not proven that this representation coincides with multiple 
polylogarithms.} have been analyzed \cite{1-llop:epsilon,korner}.  
Many physically interesting particular 
cases have been considered beyond one loop.  
Among these are the $\ep$ expansion of massless propagator 
diagrams \cite{allorder:massless} and the sunset diagram 
\cite{sunset:structure}.

\bigskip
\noindent{\bf 2. Hypergeometric Functions.}
Let us recall that there are several different ways to describe special 
functions: 
(i)  as an integral of the Euler or Mellin-Barnes type;
(ii) by a series whose coefficients satisfy certain recurrence relations; 
(iii) as a solution of a system of differential and/or difference equations (holonomic approach). 
These approaches and interrelations between them have been discussed in 
series of a papers \cite{gelfand}. In this section, we review some essential  
definitions relevant for each of these representations. 
\begin{itemize}
\item
{\bf Integral representation:} An Euler integral has the form
\begin{eqnarray}
\Phi(\vec{\alpha},\vec{\beta},P)
= 
\int_\Sigma \Pi_i P_i(x_1,\cdots, x_k)^{\beta_i} x_1^{\alpha_1} \cdots x_k^{\alpha_k} dx_1 \cdots dx_k \;,
\end{eqnarray}
where $P_i$ is some Laurent polynomial with respect to variables 
$x_1, \cdots, x_k$: 
$P_i(x_1,\cdots, x_k) = \sum c_{\omega_1 \cdots \omega_k} x_1^{\omega_1} \ldots  x_k^{\omega_k}$, 
with $\omega_j \in \mathbb{Z}$, 
and 
$
\alpha_i, \beta_j \in \mathbb{C}.
$
We assume that the region $\Sigma$ is 
chosen such that the integral exists. 

A Mellin-Barnes integral has the form 
\begin{eqnarray}
\Phi\left(a_{js},b_{kr},c_i,d_j,\gamma,\vec{x} \right) = 
\int_{\gamma+i\mathbb{R}} dz_1 \ldots dz_m
\frac{\Pi_{j=1}^{p} \Gamma\left(\sum_{s=1}^m a_{js} z_s+c_j\right)}
     {\Pi_{k=1}^{q} \Gamma\left(\sum_{r=1}^m b_{kr} z_r+d_k\right)}
x_1^{-z_1} \ldots x_m^{-z_m} \; , 
\end{eqnarray}
where 
$
a_{js}, b_{kr}, c_i, d_j \in \mathbb{R},\ \alpha_k \in \mathbb{C},
$
and $\gamma$ is chosen such that the integral exists. 
This integral can be expressed in terms of a sum of the residues of the integrated expression. 
\item
{\bf Series representation:} 
We will take the Horn definition of the series representation. 
In accordance with this definition, a formal (Laurent) power series in $r$ variables,
\begin{eqnarray}
\Phi(\vec{x}) = \sum C(\vec{m}) \vec{x}^m \equiv
\sum_{m_1,m_2,\cdots, m_r} C(m_1,m_2,\cdots,m_r) x_1^{m_1} \cdots x_r^{m_r},
\end{eqnarray}
is called {\it hypergeometric} if for each $i=1, \cdots, r$ the ratio 
$C(\vec{m}+\vec{e}_i)/C(\vec{m})$ 
is a rational function\footnote{A ``rational function'' is any function which can be written as the ratio of two polynomial functions.}
in the index of summation: $\vec{m}= (m_1, \cdots, m_r)$,
where 
$
\vec{e}_j = (0,\cdots,0,1,0,\cdots,0), 
$
is unit vector with unity in the $j^{\rm th}$ place.
Ore and Sato \cite{Ore:Sato} found that the coefficients 
of such a series have the general form 
\begin{eqnarray}
C(\vec{m})
= \Pi_{i=1}^r \lambda_i^{m_i}R(\vec{m})\Biggl( 
\Pi_{j=1}^N \Gamma(\mu_j(\vec{m})+\gamma_j+1)
\Biggr)^{-1} \;,
\end{eqnarray}
where 
$
N \geq 0, 
$
$
\lambda_j,\gamma_j \in \mathbb{C} 
$
are arbitrary complex numbers, 
$\mu_j: \mathbb{Z}^r \to \mathbb{Z}$ are arbitrary linear maps,  
and $R$ is an arbitrary rational function. 
The fact that all the $\Gamma$ factors are in the denominator is inessential: 
using the relation $\Gamma(z)\Gamma(1-z) = \pi/\sin (\pi z)$, they can be 
converted to factors in the numerator. 
A series of this type is called a ``Horn-type'' hypergeometric series. 
In this case, the system of differential equations 
has the form 
\begin{equation}
Q_j\left( 
\sum_{k=1}^r x_k\frac{\partial}{\partial x_k}
\right)
\frac{1}{x_j} \Phi(\vec{x})
= 
P_j\left( 
\sum_{k=1}^r x_k\frac{\partial}{\partial x_k}
\right)
\Phi(\vec{x}) \;, \quad j=1, \cdots, r,
\label{diff}
\end{equation}
where $P_j$ and $Q_r$ are polynomials satisfying
\begin{equation}
\frac{C(\vec{m}+e_j)}{C(\vec{m})}  =  \frac{P_j(\vec{m})}{Q_j(\vec{m})}.
\label{pre-diff}
\end{equation}
\item
{\bf Holonomic representation:} 
A combination of differential and difference equations can be found to describe
functions of the form
\begin{eqnarray}
\Phi(\vec{z},\vec{x},W) = \sum_{k_1,\cdots, k_r=0}^\infty 
\left( 
\Pi_{a=1}^m \frac{1}{z_a+\sum_{b=1}^r W_{ab} k_j}
\right) 
\Pi_{j=1}^r \frac{x_j^{k_j}}{k_j!}
\;,
\label{N}
\end{eqnarray}
where $W$ is an $r \times m$ matrix. In particular, this function satisfies the equations
\begin{eqnarray}
&& 
\frac{\partial \Phi(\vec{z},\vec{x},W)}{\partial x_j} = \Phi(\vec{z}+\omega_j,x,W) \;, \quad j=1, \cdots, r,
\\ && 
\frac{\partial}{\partial z_i}
\left( 
z_i \Phi + \sum_{j=1}^r W_{i} x_j \frac{\partial \Phi}{\partial x_j}
\right) = 0 \;, \quad i=1, \cdots, m,
\end{eqnarray}
where $\omega_j$ is the $j^{\rm th}$ column of the matrix $W$.
\end{itemize}

\bigskip
\noindent{\bf 3. Construction of the all-order $\ep$ expansion of hypergeometric functions.}
Recently, several theorems have been proven 
on the all-order $\ep$ expansion of hypergeometric 
functions about integer and/or rational values of parameters
\cite{KK08a,DK,nested1,nested2,Gauss,KWY07a,KWY07b,KWY07c}. 
For hypergeometric functions of one variable, all 
three of the representations (i)--(iii) described in the previous section are
equivalent, but some properties of the function may be more evident
in one representation than another.  

In the Euler integral representation, 
the most important results are related to the construction of the all-order 
$\ep$ expansion of Gauss hypergeometric function with special values of 
parameters in terms of Nielsen polylogarithms \cite{DK}.  
There are several important master integrals expressible in terms of this 
type of hypergeometric 
function, including one-loop propagator-type diagrams with arbitrary values 
of mass and momentum \cite{BD},
two-loop bubble diagrams with arbitrary values of masses, and 
one-loop massless vertex diagrams with three non-zero external momenta \cite{DT}.

The series representation (ii) is an intensively studied
approach. The first results of this type were derived in the context of 
the so-called ``single-scale'' diagrams \cite{series:1} related to
multiple harmonic sums.  These results have been extended to the case of 
multiple (inverse) binomial sums \cite{series:2} that correspond to the
$\ep$-expansion of hypergeometric functions with one unbalanced half-integer  
parameter and values of argument equal to $1/4$, or diagrams with two massive-particle cuts.
Particularly impressive results involving series representations were 
derived in the framework of the nested-sum approach for 
hypergeometric functions with a balanced set of parameters in 
Refs.\ \cite{nested1,nested2},
\footnote{Computer realizations of nested sums approach to expansion of hypergeometric functions 
are given in \cite{alternative,zero}.}  
and in framework of the 
generating-function approach for hypergeometric functions 
with one unbalanced set of parameters in  
Refs.\ \cite{KK08a,KWY07b,series:3,DK04}.

An approach using the iterated solution of differential equations
has been explored in Refs.\ \cite{KK08a,Gauss,KWY07a,KWY07c}.
One of the advantages of the iterated-solution approach over the series 
approach is that it provides a more efficient way to calculate each order of 
the $\ep$ expansion, since it relates each new 
term to the previously derived terms, rather than having to work with an 
increasingly large collection of independent sums at each order.
This technique includes two steps:
(i) the differential-reduction algorithm
(to reduce a generalized hypergeometric function to basic functions); 
(ii) iterative solution of the proper differential equation for the basic functions
(equivalent to iterative algorithms for calculating the analytical
coefficients of the $\ep$ expansion).

An important tool for constructing the iterative solution is the 
iterated integral defined as 
$
I(z;a_k, a_{k-1},\ldots , a_1) 
= 
\int_0^{z} \frac{dt}{t-a_k}
I(t;a_{k-1},\ldots , a_1) \;,
$
where we assume that all $a_j \neq 0$.
A special case of this integral,
$$
G_{m_k,m_{k-1},\ldots , m_1}(z;a_k, \ldots ,a_1)
\equiv
I(z;\underbrace{0, \ldots , 0}_{m_k-1 \mbox{ times}}, a_k, 
  \underbrace{0, \ldots , 0}_{m_{k-1}-1 \mbox{ times}}, a_{k-1}, 
\cdots, 
\underbrace{0, \ldots , 0}_{m_1-1 \mbox{ times}}, a_1) \;,
$$
where all $a_k \neq 0$, is related to the multiple polylogarithm \cite{Goncharov,mpl}
\begin{equation}
\Li{k_1,k_2, \ldots, k_n}{x_1, x_2, \ldots, x_n} = 
\sum_{m_n > m_{n-1} > \cdots> m_2 > m_1 > 0}^\infty \frac{x_1^{m_1}}{m_1^{k_1}} \frac{x_2^{m_2}}{m_2^{k_2}} 
\times\cdots\times \frac{x_n^{m_n}}{m_n^{k_n}}
\label{MP}
\end{equation}
by
\begin{eqnarray}
&& \hspace{-5mm}
G_{m_n,m_{n-1}, \ldots, m_1}\left(z;x_n, x_{n-1}, \ldots, x_1 \right)
 =  
(-1)^n 
\Li{m_1,m_2, \ldots, m_{n-1}, m_n}{\frac{x_2}{x_1}, \frac{x_3}{x_2}, \ldots, \frac{x_{n}}{x_{n-1}}, \frac{z}{x_n}}\;,
\nonumber \\ && \hspace{-5mm}
\Li{k_1,k_2, \ldots, k_{n-1}, k_n}{y_1, y_2, \ldots, y_{n-1}, y_n} 
= 
(-1)^n 
G_{k_n,k_{n-1},\ldots, k_2, k_1}\left(1;\frac{1}{y_n}, \ldots, \frac{1}{y_n \times\cdots\times y_1} \right)\;.\nonumber
\end{eqnarray}
In Eq.~(\ref{MP}), $k=k_1+k_2+\cdots+k_n$ is called the ``weight''  and $n$
the ``depth.''
Multiple polylogarithms (\ref{MP}) are defined for $|z_n| < 1$ and $|z_i|\leq 1 (i=1,.\cdots,n\!-\!1)$
and for $|z_n|\leq 1$ if $m_n \leq 2$.
We mention also that multiple polylogarithms form two Hopf algebras. 
One is related to the integral representation, and the other one to the series.

A particular case of the multiple polylogarithm is the 
``generalized polylogarithm'' defined by 
\begin{equation}
\Li{k_1,k_2, \ldots, k_n}{z} = 
\sum_{m_n > m_{n-1} > \cdots> m_{1}  > 0}^\infty \frac{z^{m_n}}{m_1^{k_1} m_2^{k_2} \cdots m_n^{k_n}}
= 
\Li{k_1,k_2, \ldots, k_n}{1,1,\cdots,1,z} \;, 
\label{gp}
\end{equation}
where $|z| < 1$ when all $k_i \geq 1$, or  $|z|\leq 1$ 
when $k_n \leq 2$. Another particular case is
a ``multiple polylogarithm of a square root of unity,'' defined as
\begin{equation}
\Li{\left( \sigma_1, \sigma_2, \cdots, \sigma_n \atop s_1, s_2, \cdots, s_n \right)}{z} 
= 
\sum_{m_n > m_{n-1} > \cdots m_1 > 0} z^{m_n} \frac{\sigma_n^{m_n} \cdots \sigma_1^{m_1}
                                            }{m_n^{s_n} \cdots m_1^{s_1}} \;.
\label{colored}
\end{equation}
where 
$\vec{s}=(s_1, \cdots s_n)$ and $\vec{\sigma} = (\sigma_1, \cdots, \sigma_n)$
are multi-indices and 
$\sigma_k$ belongs to the set of the square roots of unity,
$\sigma_k = \pm 1$. This particular case of multiple polylogarithms 
has been analyzed in detail by Remiddi and Vermaseren \cite{RV00}\footnote{As was pointed out by Goncharov \cite{Goncharov}, 
the iterated integral as a function of the variable $z$ has been studied by 
Kummer, Poincare, and Lappo-Danilevky, and was called a {\it hyperlogarithm}. 
Goncharov \cite{Goncharov} analyzed it as a
multivalued analytical function of $a_1,\ldots,a_k,z$. From this point of view, 
only the functions considered in Ref.\ \cite{GR00} are multiple polylogarithms 
of two variables.}.

Special consideration is necessary when the last few arguments
$a_{k-j}, a_{k-j-1}, \ldots, a_{k}$
in the integral $I(z;a_1,\cdots,a_k)$ are equal to zero, which
is called the ``trailing-zero'' case. 
It is possible to factorize such a function into a product of a power
of a logarithm and a multiple polylogarithm.
An appropriate procedure 
for multiple polylogarithms of a square root of unity was described 
in Ref.\  \cite{RV00} and extended to the case of
multiple polylogarithms in Ref.\ \cite{Weinzierl:numerical}.
For the numerical evaluation of multiple polylogarithms or its particular 
cases, see Ref.\ \cite{Weinzierl:numerical,mp:num}.
Let us consider the Laurent expansion of a generalized hypergeometric functions
of one variable
${}_{p}F_{p-1}(\vec{A};\vec{B}; z)$ with respect to its parameters. 
Such an expansion can be written as 
\begin{eqnarray}
&& 
{}_{p}F_{p-1}(\vec{A};\vec{B}; z)
 = 
{}_{p}F_{p-1}(\vec{A_0};\vec{B_0}; z)
\nonumber \\ && \hspace{-5mm}
+ 
\sum_{m_i,l_j=1}^\infty
\Pi_{i=1}^p \Pi_{j=1}^{p-1} \frac{(A_i\!-\!A_{0i})^{m_i}}{m_i!} \frac{(B_j\!-\!B_{0j})^{l_j}}{l_j!}
\left.
\left( \frac{\partial}{\partial A_i}\right)^{m_i}
\left( \frac{\partial}{\partial B_j}\right)^{l_j}
{}_{p}F_{p-1}(\vec{A};\vec{B};z)
\right|_{
\begin{smallmatrix}
A_i=A_{0i} \\
B_j=B_{0j}
\end{smallmatrix}  
} 
\nonumber \\ && \hspace{-5mm}
= 
{}_{p}F_{p-1}(\vec{A_0};\vec{B_0}; z)
+ 
\sum_{m_i,l_j=1} 
\Pi_{i=1}^p \Pi_{j=1}^{p-1} (A_i-A_{0i})^{m_i} (B_j-B_{0j})^{l_j} L_{\vec{A},{\vec{B}}}(z)
\;,
\label{Laurent}
\end{eqnarray}
where ${}_{p}F_{p-1}(\vec{A};\vec{B}; z)$ is a hypergeometric function 
defined by 
$
{}_{p}F_{p-1}(\vec{A};\vec{B}; z)\!=\!\sum_{j=0}^\infty \frac{\Pi_{i=1}^{p} (A_i)_j}{\Pi_{k=1}^{p-1}(B_k)_j} \frac{z^j}{j!} \;
$
and $(A)_j$ is the Pochhammer symbol: $(A)_j = {\Gamma(A+j)}/{\Gamma(A)}$.
Our goal is to completely describe the coefficients 
$L_{\vec{A},{\vec{B}}}(z)$ entering the r.h.s. of Eq.~(\ref{Laurent}). 
To reach this goal, we must first characterize the complete set of 
parameters for which known special functions suffice to express the 
coefficients. Beyond this, we wish to 
identity the complete set of new functions which must be invented in order to 
express all of the coefficients in the Laurent expansion.  

The first simplification comes from the well-known fact that any 
hypergeometric function 
${}_{p}F_{p-1}(\vec{A}+\vec{m};\vec{B}+\vec{k}; z)$ may be expressed
in terms of $p$ other functions of the same type:
\begin{eqnarray}
&& \hspace{-5mm}
R_{p+1}(\vec{A},\vec{B},z) {}_{p}F_{p-1}(\vec{A}+\vec{m};\vec{B}+\vec{k}; z) = 
\sum_{j=1}^{p}R_j(\vec{A},\vec{B},z) {}_{p}F_{p-1}(\vec{A}+\vec{e_k};\vec{B}+\vec{E_k}; z) \;,
\label{decomposition1}
\end{eqnarray}
where $\vec{m}, \vec{k}, \vec{e}_k$, and $\vec{E}_k$ are lists of integers, 
and the $R_k$ are polynomials in the parameters $\vec{A}, \vec{B}$, and $z$.
In particular, we can take the function and its first $p\!-\!1$ derivatives as a 
basis for the reduction (see Ref.\ \cite{D} for the details of this approach).
Then Eq.~(\ref{decomposition1}) will take the form\footnote{For simplicity,
 we will assume that no difference $B_k-A_j$ is a positive integer.}
\begin{eqnarray}
&& \hspace{-5mm}
\widetilde{R}_{p+1}(\vec{A},\vec{B},z) {}_{p}F_{p-1}(\vec{A}+\vec{m};\vec{B}+\vec{k}; z) = 
\sum_{k=1}^{p} \widetilde{R}_k(\vec{A},\vec{B},z) \left(\frac{d}{dz} \right)^{k-1} {}_{p}F_{p-1}(\vec{A};\vec{B}; z) \;,
\label{decomposition2}
\end{eqnarray}
with a new polynomial $\widetilde{R}_k$. In this way, the problem of finding the 
Laurent expansion of the original hypergeometric function
is reduced to the analysis of a set of basic functions 
and the Laurent expansion of a (formally) known polynomial. 

As is well known, hypergeometric functions satisfy the differential equation\footnote{
This equation follows from Eqs.~(\ref{diff}) -- (\ref{pre-diff}), where
$P(j) = \Pi_{k=1}^{p}(A_k+j)$ and $Q(j) = (j+1)\Pi_{k=1}^{p-1}(B_k+j)$.}
\begin{eqnarray}
\left[
z \Pi_{i=1}^{p }\left( z \frac{d}{dz} \!+\! A_i \right) 
\!-\!  z \frac{d}{dz} \Pi_{k=1}^{p-1} \left( z \frac{d}{dz} \!+\! B_k\!-\!1 \right) 
\right] {}_{p}F_{p-1}(\vec{A};\vec{B}; z) = 0.
\label{diff:eq}
\end{eqnarray}
Due to the analyticity of the hypergeometric function 
${}_{p}F_{p-1}(\vec{A};\vec{B}; z)$ 
with respect to its parameters $A_i,B_k$, \
the differential equation for the coefficients $L_{\vec{A},{\vec{B}}}(z)$
of the  Laurent expansion could be directly derived from Eq.~(\ref{diff:eq}) 
without any reference to the series or integral representation.  
This was the main idea of the approach developed in 
Refs.~\cite{KK08a,Gauss,KWY07a,KWY07c,KWY08a}.
An analysis of this system of equations and/or their explicit analytical 
solution gives us the analytical form of $L_{\vec{A},{\vec{B}}}(z)$.
It is convenient to introduce a new parametrization, 
$
A_i \to A_{0i}+a_i\ep, B_j \to B_{0i}+b_i\ep \;,
$
where $\ep$ is some small number, so that the Laurent expansion 
(\ref{Laurent}) takes the form of an ``$\ep$ expansion,'' 
$$
{}_{p}F_{p-1}(\vec{A}+\vec{a}\ep;\vec{B}+\vec{b}\ep; z)
 = 
{}_{p}F_{p-1}(\vec{A};\vec{B}; z)
+ \sum_{k=1}^\infty \ep^k L_{\vec{a},\vec{b},k}(z) 
\equiv 
\sum_{k=0}^\infty \ep^k L_{\vec{a},\vec{b},k}(z) 
\;,
$$
where $L_{\vec{a},\vec{b},0}(z) = {}_{p}F_{p-1}(\vec{A};\vec{B}; z)$. 
The differential operator can also be expanded in powers of $\ep$:
\begin{eqnarray}
D^{(p)} = 
\left[
\Pi_{i=1}^{p }\left( \theta \!+\! A_i \!+\! a_i \ep\right) 
\!-\!  \frac{1}{z} \theta \Pi_{k=1}^{p-1} \left( \theta \!+\! B_k\!-\!1 \!+\! b_k\ep\right) 
\right] = 
\sum_{j=0}^{p} \ep^j D_j^{(p-j)}(\vec{A},\vec{B},\vec{a},\vec{b},z)  \;, 
\end{eqnarray}
where 
$
\theta = z \frac{d}{d z} \;, 
$
the upper index gives the order of the differential operator, 
$
D_p^{(0)} = \Pi_{k=1}^p a_k \;,
$
 and 
\begin{eqnarray}
D_0^{(p)} & = &   
\Pi_{i=1}^{p }\left( \theta \!+\! A_i \right) 
\!-\!  \frac{1}{z} \theta \Pi_{k=1}^{p-1} \left( \theta \!+\! B_k\!-\!1 \right) 
\\ 
 & = &  \left\{ -(1\!-\!z) \frac{d}{dz}   
\!+\! \sum_{k=1}^p A_k \!-\! \frac{1}{z} \sum_{j=1}^{p-1} (B_j\!-\!1)
\right\} \theta^{p-1} 
\!+\! \sum_{j=1}^{p-1} \left[X_j(\vec{A},\vec{B}) \!-\! \frac{1}{z} Y_j(\vec{A},\vec{B}) \right] \theta^{p\!-\!1\!-\!j} \;, 
\nonumber  
\label{D0}
\end{eqnarray}
where $X_j(\vec{A},\vec{B})$ and $Y_j(\vec{A},\vec{B})$ are polynomials. 
Combining all of the expansions together, we obtain a system of 
equations
$
\sum_{r=0}^\infty \ep^r \sum_{j=0}^p  D_j^{(p-j)} L_{\vec{a},\vec{b},r-j}(z) = 0 \; , 
$
which could be split into following system (each order of $\ep$):
$(\ep^0)~D_0^{(p)} L_{\vec{a},\vec{b},0}(z) = 0 \; ;$ 
$(\ep^k, 1 \leq k \leq p)~\sum_{r=0}^k D_k^{(p-k)} L_{\vec{a},\vec{b},k-r}(z) = 0 \; ; 
$
$(\ep^k, k \geq p+1)~\sum_{j=0}^p D_j^{(p-j)} L_{\vec{a},\vec{b},k-j}(z) = 0 \;.
$
Further simplification comes from the explicit forms 
of $D_k^{(p-k)}$ and the polynomials $X_j(\vec{A},\vec{B}),Y_j(\vec{A},\vec{B})$ 
in Eq.~(\ref{D0}).  For example, for integer values of parameters, we can put 
$A_k=0,B_j=1$, so that all of the $X_j(\vec{A},\vec{B})$ and 
$Y_j(\vec{A},\vec{B})$ are equal to zero. 
Further details can be found in our papers, Refs.\ \cite{KK08a,KWY07a,KWY07b,KWY07c,KWY08a}.

Here, we will mention some of the existing results. ~\footnote{
In the following, we will assume that $a,b,c$ are an arbitrary numbers and 
$\ep$ is a small parameter.}
\begin{itemize}
\item
If $I_1,I_2,I_3$ are arbitrary integers, 
the Laurent expansions of the Gauss hypergeometric functions 
\begin{eqnarray}
&& 
{}_2F_{1}(I_1+a\ep, I_2+b\ep; I_3+\tfrac{p}{q}+c \ep;z) \;, 
\quad 
{}_2F_{1}(I_1+\tfrac{p}{q}+a\ep, I_2+\tfrac{p}{q}+b\ep; I_3+\tfrac{p}{q} + c \ep;z) \;,
\nonumber \\ &&
{}_2F_{1}(I_1+\tfrac{p}{q}+a\ep,   I_2+b\ep; I_3+c \ep;z) \;,
\quad 
{}_2F_{1}(I_1+\tfrac{p}{q}+a\ep,   I_2+b\ep; I_3+\tfrac{p}{q} + c \ep;z) 
\nonumber
\end{eqnarray}
are expressible in terms of multiple polylogarithms of arguments being powers 
of  $q$-roots of unity and a new variable, that is an algebraic function of $z$,
with coefficients that are ratios of polynomials.
\item
If $\vec{A}, \vec{B}$ are lists of integers and $I, p, q$ are integers, the 
Laurent expansions of the generalized hypergeometric functions
$$
{}_pF_{p-1}(\vec{A}+\vec{a}\ep, \tfrac{p}{q} + I; \vec{B}+\vec{b}\ep;z) \;,
\quad  
{}_pF_{p-1}(\vec{A}+\vec{a}\ep; \vec{B}+\vec{b}\ep, \tfrac{p}{q} + I;z) 
$$
are expressible in terms of multiple polylogarithms of arguments that are 
powers of $q$-roots of unity and a new variable that is an algebraic function of $z$,
with coefficients that are ratios of polynomials.
\item
If $\vec{A}, \vec{B}$ are lists of integers,
the Laurent expansion of the generalized hypergeometric function
$$
{}_pF_{p-1}(\vec{A}+\vec{a}\ep; \vec{B}+\vec{b}\ep;z) 
$$
are expressible via generalized polylogarithms (\ref{gp}).
\end{itemize}

We should also mention the following case \cite{nested2} in which the $\ep$ expansion 
has been constructed via the nested sum approach: \\
If $p, q, I_k$  are any integers and $\vec{A}, \vec{B}$ are lists of 
integers, the generalized hypergeometric function
$$
{}_pF_{p-1}(
\{\tfrac{p}{q} \!+\! \vec{A} \!+\! \vec{a}\ep\}_r, \vec{I_1} \!+\! \vec{c}\ep; 
\{\tfrac{p}{q} \!+\! \vec{B} \!+\! \vec{b}\ep\}_r, \vec{I_2} \!+\! \vec{d}\ep;z) \;
$$
is expressible in terms of multiple polylogarithms of arguments that are 
powers of $q$-roots of unity and the new variable $z^{1/q}$,
with coefficients that are ratios of polynomials.
A hypergeometric function of this form is said to 
have a {\it zero-balance} set of parameters.

We will now demonstrate some algebraic relations between functions 
generated by the $\ep$ expansion of hypergeometric functions with special sets
 of parameters.  Let us consider the analytic continuation of 
the generalized hypergeometric function $~_{p+1}F_p$
with respect to the transformation $z \to {1}/{z}$ \cite{bateman}:
\begin{eqnarray}
&& 
\left( \Pi_{j=1}^p \frac{1}{\Gamma(b_j) } \right)
~_{p+1}F_p\left(\begin{array}{c|}
a_1, a_2, \cdots, a_{p+1} \\
b_1, b_2, \cdots, b_p  \end{array} ~ z \right)  = 
\sum_{k=1}^{p+1} 
\frac{\Pi_{j=1, j \neq k}^{p+1} \Gamma(a_j \!-\! a_k)}
     {\left( \Pi_{j=1, j \neq k}^{p+1} \Gamma(a_j) \right) \left( \Pi_{j=1}^{p} \Gamma(b_j \!-\! a_k) \right)}
\nonumber \\ && \hspace{5mm}
\times
(-z)^{-a_k}
~_{p+2}F_{p+1}\left(\begin{array}{c|}
1, a_k, 1 \!+\! a_k \!-\! b_1, 1 \!+\! a_k \!-\! b_2, \cdots, 1 \!+\! a_k \!-\! b_p \\
1 \!+\! a_k \!-\! a_1, 1 \!+\! a_k \!-\! a_2, \cdots, 1 \!+\! a_k \!-\! a_{p+1}  
\end{array} ~ \frac{1}{z} \right)  
\;, 
\label{inverse}
\end{eqnarray}
where none of the differences between pairs of parameters $a_j - a_k$ 
is an integer.

On the r.h.s. of Eq.~(\ref{inverse}), we actually have a hypergeometric 
function $~_{p+1}F_p$, since one of the parameters is always equal to unity. 
If we make the replacements
$$
a_j \to \frac{r}{q} + a_j \ep \;, \quad 
b_j \to \frac{r}{q} + b_j \ep 
$$
in Eq.~(\ref{inverse}), we obtain the relation 
\begin{eqnarray}
~_{p+1}F_p\left(\begin{array}{c|}
\left\{ \frac{r}{q}+a_j \ep \right\}_{p+1} \\
\left\{ \frac{r}{q}+b_j \ep \right\}_p \end{array} ~ z \right)  
=  
\sum_{s=1}^p c_s 
~_{p+1}F_p\left(\begin{array}{c|}
\frac{r}{q} + \tilde{c} \ep, 
\left\{ 1+\tilde{a}_j \ep \right\}_{p} \\
\left\{ 1+\tilde{b}_j \ep \right\}_p \end{array} ~ \frac{1}{z} \right)  \;, 
\label{full->integer}
\end{eqnarray}
where the $c_r$ are constants. 
Another relation follows if we choose in Eq.~(\ref{inverse}) the following set of parameters: 
$$
a_j \to a_j \ep \;,  \quad j=1, \cdots,  p+1 \;, \quad 
b_k \to b_k \ep \;,  \quad k=1, \cdots,  p-1 \;, \quad 
b_{p} = \frac{r}{q} + b_p \ep \;.
$$
Then we have
\begin{eqnarray}
~_{p+1}F_p\left(\begin{array}{c|}
\left\{a_j \ep \right\}_{p+1} \\
\left\{b_j \ep \right\}_{p-1},  
\frac{r}{q} + b_p \ep 
\end{array} ~ z \right)  
= 
\sum_{s=1}^p \tilde{c}_s 
~_{p+1}F_p\left(\begin{array}{c|}
1-\frac{r}{q} - \tilde{c} \ep, 
\left\{ 1+\tilde{a}_j \ep \right\}_{p} \\
\left\{ 1+\tilde{b}_j \ep \right\}_p \end{array} ~ \frac{1}{z} \right)  \;, 
\label{down->up}
\end{eqnarray}
where the $\tilde{c}$ are constants. 
In this way, we find a proof of the following result:\\
{\bf Lemma: } {\it 
When none of the difference between two upper parameters is an integer, 
and the differences between any lower and upper parameters are 
positive integers, the 
coefficients of the $\ep$ expansion of the hypergeometric functions
$$
~_{p+1}F_p\left(\begin{array}{l|}
\vec{A}\!+\!\tfrac{r}{q}\!+\!\vec{a} \ep \\
\vec{B}\!+\!\tfrac{r}{q}\!+\!\vec{b} \ep \end{array} ~z \right)  \;, 
~_{p+1}F_p\left(\begin{array}{c|}
\vec{A}\!+\!\vec{a} \ep \\
\vec{B}\!+\!\vec{b} \ep, 
I \!+\! \tfrac{r}{q} \!+\! c \ep 
\end{array} ~z \right)  \;, 
~_{p+1}F_p\left(\begin{array}{c|}
I\!+\!\tfrac{r}{q} \!+\! c \ep, 
\vec{A}\!+\!\vec{a} \ep \\
\vec{B}\!+\!\vec{b} \ep \end{array} ~z\right)  \;, 
$$
where
$\vec{A},\vec{B},\vec{a},\vec{b}, c$ and $I$ are all integers, 
are related to each other. 
} \\ 
Note that none of the functions of this lemma belongs to the zero-balance case.

{\bf 4. One-loop vertex as hypergeometric function.}
Let us consider now the one-loop vertex diagram.
We recall that any one-loop vertex diagram with the arbitrary 
masses, external momenta and power of propagators can be reduced by 
recurrence relations to a vertex-type master integral (with all powers of propagators being equal to unity)
or, in the case of zero Gram and/or Cayley determinants, to a linear combination of 
propagator-type diagrams \cite{one-loop:reduction}. 
In the case of non-zero Gram and/or Cayley determinants, the one-loop 
master integrals are expressible in terms of linear combinations of 
two Gauss hypergeometric functions and the Appell function $F_1$ 
\cite{1loop:vertex,FJT03}.

\begin{figure}[th]
\begin{center}
\centerline{\vbox{\epsfysize=120mm \epsfbox{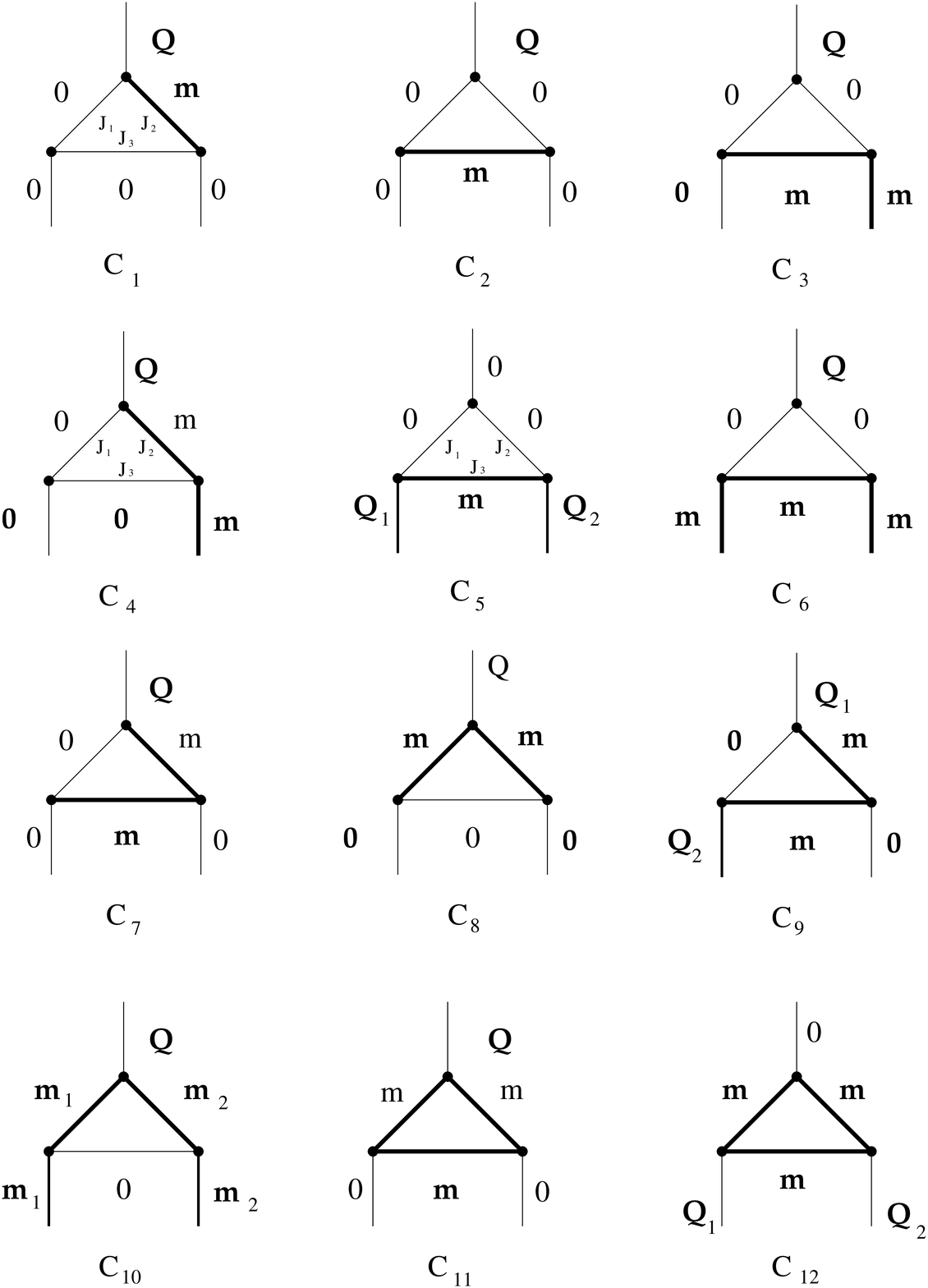}}}
\caption{\label{one-loop} \small
One-loop vertex-type diagrams expressible in terms of 
generalized hypergeometric functions.
Bold and thin lines correspond to massive and
massless propagators, respectively.}
\end{center}
\end{figure}

Our starting point is the hypergeometric representation for one-loop diagrams with three arbitrary 
external momenta and one massive line or two or three massive lines with an equal masses, derived in Ref.\ \cite{BD}.
Let us consider a one-loop vertex-type diagram, as shown in Fig.~\ref{one-loop}. 
Using properties of functions of several variables \cite{bateman,DOS}, 
these diagrams can be expressed in terms of 
hypergeometric functions of one variable\footnote{We are indebted to A.~Davydychev for assistance in that analysis.}, 
whose $\ep$ expansions up to weight 4 are presented in Ref.\ \cite{zero,DK04,FKV}
\footnote{This is enough for the calculation of two-loop corrections.}
and available via the web \cite{MKL}.
We recall that up to weight 4,
the $\ep$ expansions of all master integrals collected here are expressible in terms of Nielsen 
polylogarithms only.
The hypergeometric representations have been derived 
also in \cite{AGO2} for $C_1$ and $C_2$,
in \cite{FJT03,DOS} for  $C_6$ and 
in \cite{BD} for $C_{11}$.
Up to the finite part, some of these diagrams have been studied in 
\cite{one-loop:vertex}. 
For certain diagrams ($C_4,C_6,C_9,C_{10},C_{11}$), 
the $\ep$ expansion of the first several coefficients was given 
in Ref.\ \cite{korner} in terms of multiple polylogarithms of two variables.
We use the notations $j_{123}\!=\!j_1\!+\!j_2\!+\!j_3$, 
$j_{mn}\!=\!j_m\!+\!j_n$ below.

We will conclude with a review of the results for special cases: 

\begin{itemize}
\item
The massless triangle diagram with one massless external on-shell 
momentum  is expressible in terms of two Gauss hypergeometric functions. 
This result follows directly from a relation in Ref.\ \cite{BD}. The Cayley determinant vanishes in this case.

\item
The analytical result for diagram $C_1$ with arbitrary powers of the 
propagators is expressible 
in terms of a Gauss hypergeometric function with one integer upper parameter:
$$
\frac{C_1}{i^{1-n} \pi^{n/2}} = 
(-m^2)^{n/2\!-\!j_{123}}
\frac{\Gamma\left( j_{123} \!-\! \frac{n}{2} \right) \Gamma\left( \frac{n}{2} \!-\! j_{13} \right)}
     { \Gamma\left( \frac{n}{2} \right)  \Gamma\left( j_2 \right)}
\;{}_{2}F_1\left(\begin{array}{c|}
j_{123} \!-\! \tfrac{n}{2}, j_1 \\
\tfrac{n}{2}  \end{array} ~ \frac{Q^2}{m^2} \right) \;.
$$
The differential reduction will result in one Gauss hypergeometric function. 
The Cayley determinant vanishes for $C_1$. 

\item
The diagram $C_2$ with arbitrary powers of propagators is expressible 
in terms of two hypergeometric functions ${}_3F_2$.
In this case, both the Gram and Cayley determinants are nonzero, and the 
master integral is
\begin{eqnarray}
&& 
\frac{C_2}{i \pi^{{n}/{2}}} =
-(m^2)^{\tfrac{n}{2}-3}
\Biggl\{
\frac{\Gamma\left( 3 \!-\! \frac{n}{2} \right) \Gamma\left( \frac{n}{2} \!-\! 2 \right)}
     { \Gamma\left( \frac{n}{2} \right)}
\;{}_{2}F_1\left(\begin{array}{c|}
1, 1 \\
\tfrac{n}{2}  \end{array} ~-\frac{Q^2}{m^2} \right) 
\nonumber \\ && \hspace{5mm}
+ \left( - \frac{Q^2}{m^2}\right)^{\tfrac{n}{2}-2}
\frac{\Gamma^2\left( \frac{n}{2} \!-\! 1\right) \Gamma\left( 2 \!-\! \frac{n}{2} \right)}
     { \Gamma\left( n \!-\! 2 \right)}
\;{}_{2}F_1\left(\begin{array}{c|}
1, \tfrac{n}{2}-1 \\
n-2 \end{array} ~-\frac{Q^2}{m^2} \right) 
\Biggr\} \;.
\nonumber 
\end{eqnarray}
\item
For diagram $C_3$, the result for arbitrary powers of propagators 
is expressible in terms of the function ${}_3F_2$.
Both the Gram and Cayley determinants are nonzero,
and the master integral is a combination of two Gauss hypergeometric functions:
\begin{eqnarray}
&& 
\frac{C_3}{i \pi^{{n}/{2}}} =
-(m^2)^{\tfrac{n}{2}-3}
\frac{\Gamma\left( \frac{n}{2} \!-\! 2 \right)}
     { \Gamma\left( n \!-\! 3 \right)}
\Biggl\{
\frac{\Gamma\left( n \!-\! 4 \right)}
     { \Gamma\left( \frac{n}{2} \!-\! 1 \right)}
\;{}_{2}F_1\left(\begin{array}{c|}
1, 1 \\
5 - n  \end{array} ~\frac{Q^2}{m^2} \right) 
\nonumber \\ && \hspace{5mm}
+ \left( - \frac{Q^2}{m^2}\right)^{\tfrac{n}{2}-2}
\frac{\Gamma\left( \frac{n}{2} \!-\! 1\right) \Gamma\left( 2 \!-\! \frac{n}{2} \right)}
     { \Gamma\left( 3 \!-\! \frac{n}{2} \right)}
{\;}_{2}F_1\left(\begin{array}{c|}
1, \tfrac{n}{2}-1 \\
3-\tfrac{n}{2} \end{array} ~\frac{Q^2}{m^2} \right) 
\Biggr\} \;.
\nonumber 
\end{eqnarray}
%
%
\item
The diagram $C_4$ with arbitrary powers of propagators is
expressible in terms of a Gauss hypergeometric function with one 
integer parameter: 
$$
\frac{C_{4}}{i^{1-n} \pi^{{n}/{2}}} = 
\frac{\Gamma\left( j_{123} \!-\! \frac{n}{2} \right) 
      \Gamma\left( \frac{n}{2} \!-\! j_{13} \right)
      \Gamma\left( n \!-\! j_{12} \!-\! 2 j_3\right)}
     {
      (-m^2)^{j_{123} \!-\! \tfrac{n}{2}}
      \Gamma\left( n \!-\! j_{123} \right)
      \Gamma\left( \frac{n}{2} \!-\! j_3 \right) \Gamma(j_2)}
\;{}_{2}F_1\left(\begin{array}{c|}
j_{123} \!-\! \tfrac{n}{2}, j_1 \\
\tfrac{n}{2} \!-\! j_{3} \end{array} ~\frac{Q^2}{m^2} \right).
$$
%
%
\item
For arbitrary powers of propagators, the diagram $C_5$ 
is expressible in terms of the Appell function $F_1$:
$$
\frac{C_{5}}{i^{1-n} \pi^{{n}/{2}}} = 
(-m^2)^{\tfrac{n}{2}\!-\!j_{123}}
\frac{\Gamma\left( j_{123} \!-\! \frac{n}{2} \right) 
      \Gamma\left( \frac{n}{2} \!-\! j_{12} \right)}
     {\Gamma\left( j_{3} \right)
      \Gamma\left( \frac{n}{2} \right) }
\;{}F_1\left(
\left.
j_{123} \!-\! \tfrac{n}{2}, j_1, j_2; \tfrac{n}{2} \right| ~\frac{Q_1^2}{m^2},  \frac{Q_2^2}{m^2} \right) 
\;.
$$
When the squared external momenta are equal, $Q_1^2=Q_2^2=Q^2$,
it reduces to the Gauss hypergeometric function:
$$
\left. 
\frac{C_{5}}{i^{1-n} \pi^{{n}/{2}}} 
\right|_{Q_1^2=Q_2^2=Q^2}
= 
(-m^2)^{\tfrac{n}{2}\!-\!j_{123}}
\frac{\Gamma\left( j_{123} \!-\! \frac{n}{2} \right) 
      \Gamma\left( \frac{n}{2} \!-\! j_{12} \right)}
     {\Gamma\left( j_{3} \right)
      \Gamma\left( \frac{n}{2} \right) }
\ {}_{2}F_1\left(\begin{array}{c|}
j_{123} \!-\! \tfrac{n}{2}, j_{12} \\
\tfrac{n}{2} \end{array} ~\frac{Q^2}{m^2} \right) 
\;.
$$
For $Q_1^2=Q_2^2$, the Gram determinant is zero, and when 
$Q_1^2=Q_2^2=m^2$, the Cayley determinant is also zero. 
%
%
\item
For $C_6$, both the Gram and Cayley determinants are nonzero, and 
\begin{eqnarray}
&& 
\frac{C_6}{i \pi^{{n}/{2}}} = 
-(m^2)^{\tfrac{n}{2}-3}
\Biggl\{
\frac{\Gamma\left( 3 \!-\! \frac{n}{2} \right) \Gamma\left( n \!-\! 5 \right)}
     { \Gamma\left( n-3 \right)}
\ {}_{2}F_1\left(\begin{array}{c|}
1, 1 \\
\tfrac{7-n}{2}  \end{array} ~\frac{Q^2}{4m^2} \right) 
\nonumber \\ && \hspace{5mm}
+ \left( - \frac{Q^2}{m^2}\right)^{\tfrac{n}{2}-2}
\frac{\Gamma^2\left( \frac{n}{2} \!-\! 1\right) \Gamma\left( 2 \!-\! \frac{n}{2} \right)}
     { \Gamma\left( n \!-\! 2 \right)}
\left( \frac{3-n}{2}\right)
\ {}_{2}F_1\left(\begin{array}{c|}
1, \tfrac{n}{2}-1 \\
\frac{3}{2} \end{array} ~\frac{Q^2}{4m^2} \right) 
\Biggr\} \;.
\nonumber 
\end{eqnarray}
\end{itemize}
%
%
\begin{itemize}
%
%
\item
The diagram $C_7$ with arbitrary powers of propagators is expressible in terms 
of the function $_3F_2$.  For this diagram, both the Gram and Cayley determinants are nonzero, and the master integral is 
$$
\frac{C_7}{i \pi^{{n}/{2}}} =
-(m^2)^{\tfrac{n}{2}-3}
\frac{\Gamma\left( \frac{n}{2} \!-\! 1 \right)
      \Gamma\left( 3 \!-\! \frac{n}{2}\right)}
     { \Gamma\left( \frac{n}{2} \right)}
\ {}_{3}F_2\left(\begin{array}{c|}
1, 1, 3 - \tfrac{n}{2} \\
\tfrac{n}{2}, 2  \end{array} ~\frac{Q^2}{m^2} \right) 
\;.
$$
%
%
%
\item
The diagram $C_8$ with arbitrary powers of propagators is expressible in 
terms of the function $_4F_3$. For this diagram, both the Gram and Cayley 
determinants are nonzero.  The master integral is 
$$
\frac{C_8}{i \pi^{{n}/{2}}} =
-(m^2)^{\tfrac{n}{2}-3}
\frac{\Gamma\left( \frac{n}{2} \!-\! 1 \right)
      \Gamma\left( 3 \!-\! \frac{n}{2}\right)}
     { \Gamma\left( \frac{n}{2} \right)}
\ {}_{3}F_2\left(\begin{array}{c|}
1, 3 - \tfrac{n}{2}, \tfrac{n}{2}-1 \\
\tfrac{n}{2}, \tfrac{3}{2}  \end{array} ~\frac{Q^2}{4m^2} \right) 
\;.
$$
%
%
%
%
%
\item
For $C_{9}$, both the Gram and Cayley determinants are nonzero.
\begin{eqnarray}
\frac{C_{9}}{i \pi^{{n}/{2}}} & = &  
-(m^2)^{\tfrac{n}{2}-3}
\frac{\Gamma\left( 3 \!-\! \frac{n}{2} \right) \Gamma\left( \frac{n}{2} \!-\! 1\right)}
     { \Gamma\left( \frac{n}{2} \right)}
\frac{1}{Q_1^2-Q_2^2}
\nonumber \\ && 
\times
\Biggl\{
{}_{3}F_2\left(\begin{array}{c|}
3 \!-\! \tfrac{n}{2}, 1, 1 \\
\tfrac{n}{2},2   \end{array} ~ \frac{Q_1^2}{m^2} \right) 
Q_1^2
- 
{}_{3}F_2\left(\begin{array}{c|}
3 \!-\! \tfrac{n}{2}, 1,1 \\
\tfrac{n}{2},2  \end{array} ~ \frac{Q_2^2}{m^2} \right) 
Q_2^2
\Biggr\} \;.
\nonumber 
\end{eqnarray}
When $Q_1^2=Q_2^2$, the Gram determinant is equal to zero. 
%
%
\item
For diagram $C_{10}$, the Cayley determinant vanishes, so that the diagram 
can be reduced to a linear combination of 
one-loop propagator diagrams (see Ref.\ \cite{DK}). 
The hypergeometric function representation is 
\begin{eqnarray}
&& 
\frac{C_{10}}{i \pi^{{n}/{2}}} = 
- \frac{\Gamma\left( 3-\frac{n}{2}\right)}{2 Q^2 (n-4)}
\nonumber \\ && \hspace{5mm}
\times
\Biggl\{
(Q^2\!+\!m_1^2\!-\!m_2^2) (m_1^2)^{\tfrac{n}{2}-3}
\ {}_{2}F_1\left(\begin{array}{c|}
1, 3 \!-\! \tfrac{n}{2} \\
\tfrac{3}{2} \end{array} ~ \frac{(Q^2+m_1^2-m_2^2)^2}{4 m_1^2 Q^2} \right) 
\nonumber \\ && \hspace{10mm}
+ 
(Q^2\!-\!m_1^2\!+\!m_2^2) (m_2^2)^{\tfrac{n}{2}-3}
\ {}_{2}F_1\left(\begin{array}{c|}
1, 3 \!-\! \tfrac{n}{2} \\
\tfrac{3}{2} \end{array} ~ \frac{(Q^2-m_1^2+m_2^2)^2}{4 m_2^2 Q^2} \right) 
\Biggr\} \;.
\nonumber 
\end{eqnarray}
%
%
%
%
%
%
%
\item
For this diagram, both the Gram and Cayley determinants 
are nonzero.  The master integral is 
$$
\frac{C_{11}}{i \pi^{{n}/{2}}} = 
-\frac{1}{2} (m^2)^{\tfrac{n}{2}\!-\!3} \Gamma\left( 3 \!-\! \frac{n}{2} \right)
\ {}_{3}F_2\left(\begin{array}{c|}
3 \!-\! \frac{n}{2}, 1, 1 \\
\frac{3}{2},  2 \end{array} ~\frac{Q^2}{4m^2} \right) 
\;.
$$
The all-order $\ep$ expansions of $C_{11}$ is 
expressible in terms of multiple polylogarithm of a square root of unity.
\item
The master integral for diagram $C_{12}$ was evaluated in Ref.\ \cite{DOS} 
in terms of a linear combination of two $_3F_2$ functions of the same type, as 
for the diagram $C_{8}$:
\begin{eqnarray}
\frac{C_{12}}{i \pi^{\tfrac{n}{2}}} & = &  
-(m^2)^{\tfrac{n}{2}-3}
\Gamma\left( 3 \!-\! \frac{n}{2} \right) 
\frac{1}{2(Q_1^2-Q_2^2)}
\nonumber \\ && 
\times
\Biggl\{
{}_{3}F_2\left(\begin{array}{c|}
3 \!-\! \tfrac{n}{2}, 1, 1 \\
\tfrac{3}{2},2   \end{array} ~ \frac{Q_1^2}{4m^2} \right) 
Q_1^2
- 
{}_{3}F_2\left(\begin{array}{c|}
3 \!-\! \tfrac{n}{2}, 1,1 \\
\tfrac{3}{2}, 2  \end{array} ~ \frac{Q_2^2}{m^2} \right) 
Q_2^2
\Biggr\} \;.
\nonumber 
\end{eqnarray}
\end{itemize}
%
%
%
%
%

\bigskip
\noindent{\bf Acknowledgments.}
\smallskip
M.Yu.K. is grateful to the Organizers of ``Quark-2008'' for their hospitality and 
to all participants, but especially to K.~Chetyrkin, A.~Isaev, A.~Kataev, S.~Larin and A.~Pivovarov,
for useful discussion. 
We are indebted to A.~Davydychev and O.~Tarasov 
for a careful reading of manuscript. 
M.Yu.K. is thankful to A.~Kotikov, T.~Huber and D.~Ma\^{\i}tre for useful comments. 
M.Yu.K.'s research was supported in part by BMBF Grant No.\ 05~HT6GUA 
and DFG Grant No.\ KN~365/3-1. 
B.F.L.W.'s research was partly supported by US DOE grant DE-FG02-05ER41399 
and by NATO grant PST.CLG.980342.



\begin{thebibliography}{99}
\bibitem{regge}
T.~Regge, {\it Algebraic Topology Methods in the Theory of Feynman Relativistic 
Amplitudes}, Battelle Rencontres, 1967. {\it Lectures in Mathematics and 
Physics}, ed. C. M. DeWitt, J. A. Wheeler.  New York: W. A. Benjamin 1968.

\bibitem{landau}
L.D.~Landau,
Nucl.\ Phys.\  {\bf 13} (1959) 181; \\
N.~Nakanishi, Prog.\ Theor.\ Phys. {\bf 22} (1959) 128; 
{\it ibid} {\bf 23} (1960) 284. 

\bibitem{book}
R.J.~Eden,  P.V.~Landshoff, D.I.~Olive, J.C.~Polkinghorne, 
{\it The Analytic $S$-Matrix}, 
Cambridge,  Cambridge University Press 1966; \\
R.~Hwa, V.~Teplitz, 
{\it Homology and Feynman Integrals},
W.A.Benjamin, New York, 1966; \\
J.D.~Bjorken, Doctoral dissertation, Stanford University, 1959.

\bibitem{1loop:series}
D.S.~Kershaw,
Phys.\ Rev.\ D {\bf 8} (1973) 2708; 
A.C.T.~Wu,
Phys.\ Rev.\ D {\bf 9} (1974) 370; \\
K.~Mano,
Phys.\ Rev.\  D {\bf 11} (1975) 452.

\bibitem{Lappo}
J.A.~Lappo-Danilevsky, 
{\it Theory of Functions on Matrices and Systems of 
Linear Differential Equations}
(Leningrad, 1934).

\bibitem{1loop:diff}
G.~Barucchi, G.~Ponzano,
J.\ Math.\ Phys.\  14 (1973) 396.

\bibitem{monodromy}
G.~Ponzano, T.~Regge, E.R.~Speer, M.J.~Westwater,
Commun.\ Math.\ Phys.\  {\bf 15} (1969) 83; 
ibid {\bf 18} (1970) 1; 
T.~Regge, E.R.~Speer, M.J.~Westwater,
Fortsch.\ Phys.\  {\bf 20} (1972) 365.

\bibitem{golubeva}
V.A.~Golubeva,
Russ.\ Math.\ Surv.\ {\bf 31} (1976) 139.

\bibitem{KK}
M.~Kashiwara, T.~Kawai,
Publ.\ Res.\ Inst.\ Math.\ Sci.\ Kyoto {\bf 12} (1977) 131;
Commun.\ Math.\ Phys.\  {\bf 54} (1977) 121;  \\
T.~Kawai, H.P.~Stapp,
Commun.\ Math.\ Phys.\  {\bf 83} (1982) 213.

\bibitem{QFT}
N.N.~Bogoliubov, D.V.~Shirkov, 
{\it Introduction to the Theory of Quantized Fields},
(Wiley \& Sons, New York, 1980); \\
C.~Itzykson, J.B.~Zuber,  {\it Quantum Field Theory}
(McGraw-Hill, New York, 1980).

\bibitem{dimreg}
G.~'tHooft, M.~Veltman,
Nucl.\ Phys.\ B {\bf 44} (1972) 189.

\bibitem{ibp}
F.V.~Tkachov, 
Phys.\ Lett.\ B {\bf 100} (1981) 65;\\
K.G.~Chetyrkin, F.V.~Tkachov, 
Nucl.\ Phys.\ B {\bf 192} (1981) 159.

\bibitem{tarasov}
O.V.~Tarasov, 
Phys.\ Rev.\ D {\bf 54} (1996) 6479.

\bibitem{Davydychev:1991}
A.I.~Davydychev,
Phys.\ Lett.\  B {\bf 263} (1991) 107.

\bibitem{rainville}
E.D.~Rainville, 
{\it Special Functions} (MacMillan Co., New York, 1960).

\bibitem{D}
M.~Saito, B.~Sturmfels, N.~Takayama, 
{\it Gr\"obner Deformations of Hypergeometric Differential Equations},
(Springer-Verlag, Berlin, 2000).

\bibitem{MKL06}
M.Yu.~Kalmykov,
JHEP {\bf 0604} (2006) 056. 

\bibitem{Tarasov}
O.V.~Tarasov,
Acta Phys.\ Polon.\  B {\bf 29} (1998) 2655.

\bibitem{DE}
A.V.~Kotikov,
Phys.\ Lett.\ B {\bf 254} (1991) 158;
{\it ibid} {\bf 259} (1991) 314;
{\it ibid} {\bf 267} (1991) 123;\\
E.~Remiddi,
Nuovo Cim.\ A {\bf 110} (1997) 1435.

\bibitem{DE:old}
G.~'t Hooft, M.J.G.~Veltman,
Nucl.\ Phys.\  B {\bf 44} (1972) 189; \\
G.~Rufa,
Annalen Phys.\ {\bf 47} (1990) 6.

\bibitem{DE:recent}
M.~Argeri, P.~Mastrolia,
Int.\ J.\ Mod.\ Phys.\  A {\bf 22} (2007) 4375.

\bibitem{ADE}
V.V~Golubev, 
{\it Lectures on the Analytic Theory of Differential Equations}, 
2$^{\rm nd}$ ed. (Gosudarstv.\ Izdat.\ Tehn.-Teor.\ Lit., Moscow-Leningrad, 
1950).

\bibitem{heun}
O.V.~Tarasov,
Phys.\ Lett.\ B {\bf 638} (2006) 195; \\
U.~Aglietti, R.~Bonciani, L.~Grassi, E.~Remiddi,
Nucl.\ Phys.\  B {\bf 789} (2008) 45.

\bibitem{num}
F.V.~Tkachov,
Nucl.\ Instrum.\ Meth.\  A {\bf 389} (1997) 309; \\
G.~Passarino,
Nucl.\ Phys.\  B {\bf 619} (2001) 257; \\
G.~Passarino, S.~Uccirati,
Nucl.\ Phys.\  B {\bf 629} (2002) 97; \\
F.~Jegerlehner, M.Yu.~Kalmykov, O.~Veretin, 
Nucl.\ Phys.\ B  {\bf 641} (2002) 285; \\
A.~Ferroglia, M.~Passera, G.~Passarino, S.~Uccirati,
Nucl.\ Phys.\  B {\bf 650} (2003) 162; \\
C.~Anastasiou, A.~Daleo,
JHEP {\bf 0610} (2006) 031; \\
C.~Anastasiou, S.~Beerli, A.~Daleo,
JHEP {\bf 0705} (2007) 071; \\
S.~Actis, G.~Passarino, C.~Sturm, S.~Uccirati,
arXiv:0809.3667;  \\
G.~Heinrich,
Int.\ J.\ Mod.\ Phys.\  A {\bf 23} (2008) 1457.

\bibitem{davydychev}
A.I.~Davydychev,
J.\ Math.\ Phys.\  {\bf 32} (1991) 1052; 
ibid {\bf 33} (1992) 358.

\bibitem{BD}
E.E.~Boos, A.I.~Davydychev,
Theor.\ Math.\ Phys.\  {\bf 89} (1991) 1052.

\bibitem{1loop:vertex}
O.V.~Tarasov,
Nucl.\ Phys.\ Proc.\ Suppl.\  {\bf 89} (2000) 237.

\bibitem{FJT03}
J.~Fleischer, F.~Jegerlehner, O.V.~Tarasov,
Nucl.\ Phys.\ B {\bf 672} (2003) 303.

\bibitem{one-loop:reduction}
J.~Fleischer, F.~Jegerlehner, O.V.~Tarasov,
Nucl.\ Phys.\ B {\bf 566} (2000) 423; \\
T.~Binoth, J.~P.~Guillet, G.~Heinrich,
Nucl.\ Phys.\ B {\bf 572} (2000) 361; \\
T.~Binoth, J.P.~Guillet, G.~Heinrich, E.~Pilon, C.~Schubert,
JHEP {\bf 0510} (2005) 015.

\bibitem{one-loop:nonexceptional}
G.~Passarino, M.J.G.~Veltman,
Nucl.\ Phys.\  B {\bf 160} (1979) 151; \\
A.V.~Kotikov,
Mod.\ Phys.\ Lett.\  A {\bf 6} (1991) 3133.

\bibitem{lauricella}
F.A.~Berends, M.~Buza, M.~B\"ohm, R.~Scharf,
Z.\ Phys.\ C {\bf 63} (1994) 227.

\bibitem{sunset}
A.I.~Davydychev,
{\em ``Loop calculations in QCD with massive quarks"},
talk at Int.\ Conf.\ ``Relativistic Nuclear Dynamics"
(Vladivostok, Russia, September 1991),\\ 
D.J.~Broadhurst, J.~Fleischer, O.V.~Tarasov,
Z.\ Phys.\ C {\bf 60} (1993) 287; \\
A.I.~Davydychev, A.G.~Grozin,
Phys.\ Rev.\ D {\bf 59} (1999) 054023; \\
F.~Jegerlehner, M.Yu.~Kalmykov,
Nucl.\ Phys.\ B {\bf 676} (2004) 365.

\bibitem{KK08a}
M.Yu.~Kalmykov, B.~Kniehl, doi: 10.1016/j.nuclphysb.2008.08.022 
(arXiv:0807.0567).

\bibitem{bateman}
A.~Erdelyi (Ed.), {\it Higher Transcendental Functions}, vol.1 (McGraw-Hill, New York, 1953); 
L.J.~Slater, {\it Generalized Hypergeometric Functions} (Cambridge University Press, Cambridge 1966).

\bibitem{one-loop:finite}
G.~'t Hooft, M.J.G.~Veltman,
Nucl.\ Phys.\  B {\bf 153} (1979) 365; \\
A.~Denner, U.~Nierste, R.~Scharf,
Nucl.\ Phys.\  B {\bf 367} (1991) 637.

\bibitem{Lewin}
L.~Lewin, {\it Polylogarithms and associated functions}
(North-Holland, Amsterdam, 1981).

\bibitem{DK}
A.I.~Davydychev,
Phys.\ Rev.\ D {\bf 61} (2000) 087701; \\
A.I.~Davydychev, M.Yu.~Kalmykov,
Nucl.\ Phys.\ Proc.\ Suppl.\  {\bf 89} (2000) 283; 
Nucl.\ Phys.\ B {\bf 605} (2001) 266; 
arXiv:hep-th/0203212.

\bibitem{one-loop:linear}
U.~Nierste, D.~M\"uller, M.~B\"ohm,
Z.\ Phys.\  C {\bf 57} (1993) 605.

\bibitem{1loop:all}
A.I.~Davydychev,
Nucl.\ Instrum.\ Meth.\ A {\bf 559} (2006) 293; 
O.V.~Tarasov,
arXiv:0809.3028.

\bibitem{Goncharov}
A.B.~Goncharov,
{\it Proceedings of the International Congress of Mathematicians, Zurich, 1994} 
(Birkh\"auser, Basel, 1995) Vol.\ 1, 2, p.~374;
Math.\ Res.\ Lett.\  {\bf 4}  (1997) 617; 
{\it ibid} {\bf 5}  (1998) 497; 
arXiv:math/0103059.

\bibitem{1-llop:epsilon}
J.~Fleischer, T.~Riemann, O.V.~Tarasov,
Acta Phys.\ Polon.\  B {\bf 34} (2003) 5345.

\bibitem{korner}
J.G.~K\"orner, Z.~Merebashvili, M.~Rogal,
Phys.\ Rev.\ D {\bf 71} (2005) 054028; 
J.\ Math.\ Phys.\  {\bf 47} (2006) 072302.

\bibitem{allorder:massless}
D.J.~Broadhurst, D.~Kreimer,
Int.\ J.\ Mod.\ Phys.\  C {\bf 6} (1995) 519; 
Phys.\ Lett.\  B {\bf 393} (1997) 403; 
I.~Bierenbaum, S.~Weinzierl,
Eur.\ Phys.\ J.\ C  {\bf 32} (2003) 67; 
F.~Brown,
arXiv:0804.1660.

\bibitem{sunset:structure}
S.~Bauberger, F.~A.~Berends, M.~Bohm, M.~Buza,
Nucl.\ Phys.\  B {\bf 434} (1995) 383; \\
A.I.~Davydychev, R.~Delbourgo,
J.\ Phys.\ A  {\bf 37} (2004) 4871; \\
G.~Passarino,
Nucl.\ Phys.\ Proc.\ Suppl.\  {\bf 135} (2004) 265; \\
B.A.~Kniehl et al.,
Nucl.\ Phys.\  B {\bf 738} (2006) 306; \\
D.H.~Bailey et al.,
J. Phys.\ A {\bf 41} (2008) 20520;  \\
S.~Laporta,
Phys.\ Lett.\  B {\bf 549} (2002) 115; 
arXiv:0803.1007; \\
P.~Aluffi, M.~Marcolli,
arXiv:0807.1690

\bibitem{gelfand}
I.M.~Gelfand, M.M.~Kapranov, A.V.~Zelevinsky, 
Adv.\ Math.\  {\bf 84} (1990) 255; \\
I.M.~Gel'fand, M.I.~Graev, V.S.~Retakh, 
Russian Math.\ Surveys  {\bf 47}  (1992) 1; \\
I.M.~Gelfand, M.I.~Graev,
Russian Math. Surveys {\bf 52} (1997) 639; 
{\it ibid} {\bf 56} (2001) 615.

\bibitem{Ore:Sato}
O.~Ore, 
J.\ Math.\ Pure Appl.\ {\bf 9} (1930) 311; 
M.~Sato, 
Nagoya\ Math.\ J.\ {\bf 120} (1990) 1.

\bibitem{nested1}
S.~Moch, P.~Uwer, S.~Weinzierl,
J.\ Math.\ Phys.\  {\bf 43} (2002) 3363.

\bibitem{nested2}
S.~Weinzierl,
J.\ Math.\ Phys.\  {\bf 45} (2004) 2656.

\bibitem{Gauss}
Shu~Oi, math.NT/0405162.

\bibitem{KWY07a}
M.Yu.~Kalmykov, B.F.L.~Ward, S.~Yost,
JHEP {\bf 0702} (2007) 040.

\bibitem{KWY07b}
M.Yu.~Kalmykov, B.F.L.~Ward, S.A.~Yost,
JHEP {\bf 0710} (2007) 048.

\bibitem{KWY07c}
M.Yu.~Kalmykov, B.F.L.~Ward,  S.A.~Yost,
JHEP {\bf 0711} (2007) 009. 

\bibitem{DT}
A.I.~Davydychev, J.B.~Tausk,
Nucl.\ Phys.\  B {\bf 397} (1993) 123; 
Phys.\ Rev.\ D {\bf 53} (1996) 7381. 

\bibitem{series:1}
N.~Gray, D.J.~Broadhurst, W.~Grafe, K.~Schilcher,
Z.\ Phys.\ C {\bf 48} (1990) 673; \\
D.J.~Broadhurst,
Z.\ Phys.\ C {\bf 47} (1990) 115;
{\it ibid} {\bf 54} (1992) 599; 
arXiv:hep-th/9604128; \\
J.M.~Borwein, D.M.~Bradley, D.J.~Broadhurst,
Electron. J. Combin. {\bf 4} (1997) \#R5; \\
J.A.M.~Vermaseren,
Int.\ J.\ Mod.\ Phys.\  A {\bf 14} (1999) 2037; \\
M.~Bigotte, G.~Jacob, N.E.~Oussous, M.~Petitot, 
Theoret.\ Comput.\ Sci.\  {\bf 273}  (2002) 271.

\bibitem{alternative}
S.~Weinzierl,
Comput.\ Phys.\ Commun.\  {\bf 145} (2002) 357; \\
S.~Moch, P.~Uwer,
Comput.\ Phys.\ Commun.\  {\bf 174} (2006) 759; \\
T.~Huber, D.~Ma\^{\i}tre,
Comput.\ Phys.\ Commun.\  {\bf 175} (2006) 122.

\bibitem{zero}
T.~Huber, D.~Ma\^{\i}tre,
Comput.\ Phys.\ Commun.\  {\bf 178} (2008) 755.

\bibitem{series:2}
D.J.~Broadhurst,
Eur.\ Phys.\ J.\ C{\bf 8} (1999) 311; \\
J.~Fleischer, M.Yu.~Kalmykov, A.V.~Kotikov,
Phys.\ Lett.\ B {\bf 462} (1999) 169; \\
J.~Fleischer, M.Yu.~Kalmykov,
Phys.\ Lett.\ B {\bf 470} (1999) 168; \\
J.M.~Borwein, D.J.~Broadhurst, J.~Kamnitzer,
Exper.\ Math.\  {\bf 10} (2001) 25; \\
M.Yu.~Kalmykov, O.~Veretin,
Phys.\ Lett.\ B {\bf 483} (2000) 315; \\
M.Yu.~Kalmykov, A.~Sheplyakov,
Comput.\ Phys.\ Commun.\  {\bf 172} (2005) 45; \\
M.Yu.~Kalmykov,
Nucl.\ Phys.\ B {\bf 718} (2005) 276; \\
Jianqiang Zhao,  
arXiv:math/0302055.

\bibitem{series:3}
F.~Jegerlehner, M.Yu.~Kalmykov, O.~Veretin, 
Nucl.\ Phys.\ B {\bf 658} (2003) 49.

\bibitem{DK04}
A.I.~Davydychev, M.Yu.~Kalmykov,
Nucl.\ Phys.\ B {\bf 699} (2004) 3; \\
M.Yu.~Kalmykov,
Nucl.\ Phys.\ Proc.\ Suppl.\  {\bf 135} (2004) 280.


\bibitem{KWY08a}
S.A.~Yost, M.Yu.~Kalmykov, B.F.L.~Ward,
ICHEP 2008, Philadelphia, arXiv:0808.2605.

\bibitem{mpl}
J.M.~Borwein et al.,
Trans.\ Am.\ Math.\ Soc.\  {\bf 353} (2001) 907; \\
M.~Waldschmidt, 
``Multiple polylogarithms: an introduction,''  in
{\it Number theory and discrete mathematics} (Chandigarh, 2000),  1--12, (Trends Math., Birkh{\"a}user, Basel, 2002).

\bibitem{RV00}
E.~Remiddi, J.A.M.~Vermaseren,
Int.\ J.\ Mod.\ Phys.\ A {\bf 15} (2000) 725.

\bibitem{GR00}
T.~Gehrmann, E.~Remiddi,
Nucl.\ Phys.\  B {\bf 601}  (2001) 248.

\bibitem{Weinzierl:numerical}
J.~Vollinga, S.~Weinzierl,
Comput.\ Phys.\ Commun.\  {\bf 167} (2005) 177.

\bibitem{mp:num}
D.~Ma\^{\i}tre,
Comput.\ Phys.\ Commun.\  {\bf 174} (2006) 222; 
arXiv:hep-ph/0703052.

\bibitem{FKV}
J.~Fleischer, A.V.~Kotikov, O.L.~Veretin,
Nucl.\ Phys.\  B {\bf 547} (1999) 343

\bibitem{DOS}
A.I.~Davydychev, P.~Osland, L.~Saks, 
Phys.\ Rev.\ D {\bf 63} (2001) 014022; 
JHEP {\bf 0108} (2001) 050.

\bibitem{AGO2}
C.~Anastasiou, E.W.N.~Glover, C.~Oleari, 
Nucl.\ Phys.\ B {\bf 572} (2000) 307.

\bibitem{one-loop:vertex}
R.K.~Ellis, G.~Zanderighi,
JHEP {\bf 0802} (2008) 002; \\
J.R.~Andersen, T.~Binoth, G.~Heinrich, J.M.~Smillie,
JHEP {\bf 0802} (2008) 057.
%
\bibitem{MKL}
M.Yu. Kalmykov, 
{\it Hypergeometric functions: reduction and epsilon-expansion},
http://theor.jinr.ru/$\;\widetilde{}\;$kalmykov/hypergeom/hyper.html


\end{thebibliography}
\end{document}